\documentclass[onecolumn,draftcls]{IEEEtran}
\normalsize
\usepackage{amsmath,amsfonts,amssymb,amsbsy,url,verbatim}
\usepackage{comment}
\usepackage{graphicx}
\usepackage{url}
\usepackage{times}
\usepackage{dsfont}
\usepackage{subfigure}
\usepackage{cite}
\usepackage{bm}
\usepackage{epstopdf}
\usepackage{algorithm,algorithmic, color}
\usepackage{amsthm}
\usepackage{bm}

\renewcommand{\vec}[1]{\bm{#1}}
\newcommand{\mat}[1]{\mathbf{#1}}
\newcommand{\hermitian}{^{H\scriptscriptstyle}}
\newcommand{\transpose}{^{T\scriptscriptstyle}}
\newcommand{\radixM}{^{\frac{1}{2}\scriptscriptstyle}}
\newcommand{\radixH}{^{\frac{H}{2}\scriptscriptstyle}}

\newcommand{\bi}{\begin{itemize}}
\newcommand{\ei}{\end{itemize}}
\newcommand{\ben}{\begin{enumerate}}
\newcommand{\een}{\end{enumerate}}
\newcommand{\bc}{\begin{cases}}
\newcommand{\ec}{\end{cases}}
\newcommand{\bd}{\begin{description}}
\newcommand{\ed}{\end{description}}

\newcommand{\be}{\begin{equation}}
\newcommand{\ee}{\end{equation}}
\newcommand{\bea}{\begin{eqnarray}}
\newcommand{\eea}{\end{eqnarray}}

\newcommand{\rank}{{\mathrm{rank}}}

\IEEEoverridecommandlockouts

\begin{document}

\title{Pilot Length Optimization for Spatially Correlated Multi-User MIMO Channel Estimation}

\author{Beatrice~Tomasi, Maxime~Guillaud\\
Mathematical and Algorithmic Sciences Laboratory, Huawei Technologies Co. Ltd.\\
France Research Center, 20 quai du Point du Jour, 92100 Boulogne Billancourt, France\\
email: \texttt{\{beatrice.tomasi,maxime.guillaud\}@huawei.com}}

\maketitle

\begin{abstract}
We address the design of pilot sequences for channel estimation in the context of multiple-user Massive MIMO; considering the presence of channel correlation, and assuming that the statistics are known, we seek to exploit the spatial correlation of the channels to  minimize the length of the pilot sequences, and specifically the fact that the users can be separated either through their spatial signature (low-rank channel covariance matrices), or through the use of different training sequences.
We introduce an algorithm to design short training sequences for a given set of user covariance matrices. The obtained pilot sequences are in general non-orthogonal, however they ensure that the channel estimation error variance is uniformly upper-bounded by a chosen constant over all channel dimensions.
We show through simulations using a realistic scenario based on the one-ring channel model that the proposed technique can yield pilot sequences of length significantly smaller than the number of users in the system.
\end{abstract}

\section{Introduction}

Channel state information (CSI) acquisition represents an important problem in the multi-user Massive MIMO (Multiple-Input Multiple-Output) scenario \cite{Larsson_MassiveMIMO_IEEE_CommMag2014}. Accurate downlink CSI is required in order to obtain the large multiplexing gain expected in massive MIMO system and achieve the rates shown e.g. in~\cite{JSAC2013HoydisDebbah}. 
It is well known that, in the presence of i.i.d. channels, it is necessary to make the length of the pilot sequences at least as large as the total number of transmit antennas, in order to avoid the effect known as pilot contamination \cite{Ngo_Marzetta_Larsson_contamination_ICASSP2011}; depending on the coherence time of the channel, the transmission of long training sequences instead of data-bearing symbols can represent a significant loss in spectral efficiency.
In this context, classical uplink CSI acquisition based on orthogonal pilot sequences across the users may not be efficient, since maintaining pilot orthogonality across many users requires the use of longer training sequences than required. 

In the context of Massive MIMO however, the channels exhibit a large degree of correlation \cite{Hoydis_MMIMO_measurements_ISWCS2012}.
In fact, a denser antenna array improves the spatial resolution, and makes the received signal more spatially correlated, to the point of resulting in a rank deficient spatial correlation matrix. This correlation can potentially help reduce the required training overhead, since the requirement to maintain pilot orthogonality across many users can be relaxed.
In the literature, the problem of pilot design for MIMO correlated channels has been widely studied, for example in~\cite{TSP2004KotechaSayeed}, \cite{TSP2010BjornsonOttersten}, and \cite{TSP2014ShariatiBengtsson}. However, in these works the pilot optimization is done for a MIMO system where all the transmitting antennas share the same spatial correlation subspace.
This assumption was lifted in \cite{Adhikary_JSDM_IT2013}, where it is proposed to schedule uplink CSI acquisition across the users such that the terminals can be separated in space; pilot orthogonality in time is therefore not required, yielding shorter training sequences. However, it is not clear how close to perfect separation in space a practical system can operate, considering realistic propagation conditions, and a finite number of users to choose from at the scheduling stage.

Another aspect of the problem, related to mitigating pilot contamination in the context of multi-cell Massive MIMO, has been considered in e.g. \cite{Yin_Gesbert_etal_coordinated_estimation_JSAC2013}; in this work, the same (orthogonal) pilot sequences are reused across the cells, and the problem of assigning each user to one of the pilot sequences is considered. 

The object of the present article is to explore the options available for the design of pilot sequences when the users' spatial covariances are not strictly orthogonal. We focus on the single-cell case, however we do not require orthogonality in time between the pilot sequences assigned to different users.  In the following, we target this problem in the context of uplink CSI acquisition, when the channel covariances of the individual users are assumed to be known and arbitrary (i.e. they can be either mutually orthogonal, or have partly or fully overlapping spans). In Section~\ref{section_identifiability}, we consider the noiseless case, and establish bounds on the length of the training sequences required to simultaneously learn all the users' channels, by deriving necessary and sufficient conditions for channel identifiability. In Section~\ref{section_optimization}, we introduce an algorithm to design pilot sequences with the objective of minimizing their length across all the users, while simultaneously ensuring that the channel estimation error variance is uniformly upper-bounded by a chosen constant over all channel dimensions. Finally, in Section~\ref{section_simulations}, we show through simulations using a realistic scenario based on the one-ring channel model that the proposed technique can yield pilot sequences of length significantly smaller than the number of users in the system.

\section{System Model}

Let us consider a massive MIMO system with $K$ single-antenna user terminals (UT) and $M$ antennas at the BS. 
The column vector of channel coefficients between the $k$-th single-antenna terminal and the $M$ antennas at the BS, $\vec{h}_k \in \mathbb{C}^{M}$ can be expressed as the product between the spatial correlation matrix $\mat{R}_k\radixM  \in \mathbb{C}^{M\times r_k}$ where $r_k \leq M$ is the rank of the spatial correlation, and a vector of complex Gaussian i.i.d. random variables, $\vec{\eta}_k \in \mathbb{C}^{r_k}$ that represents the fast fading process, i.e.
\be \label{ch_expr}
\vec{h}_k = \mat{R}_k\radixM \vec{\eta}_k, \quad \quad \forall k=1\ldots K.
\ee
We assume that the $\vec{\eta}_k \sim \mathcal{CN}(0,\mat{I}_{r_k})$ are independent across the users, and that the spatial correlation matrices $\mat{R}_k =\mathbb{E}[\vec{h}_k\vec{h}_k^H]$ are constant and known at the BS. Note that the single-antenna assumption is made here as a matter of notational simplicity, and can be trivially relaxed by treating the antennas of a multi-antenna user as multiple virtual users having the same channel covariance.

We assume that pilot sequences are sent simultaneously from all terminals to the BS in order to estimate the channel coefficients. Let $\vec{p}_k \in  \mathbb{C}^{L}$ denote the sequence of length $L$ symbols transmitted by terminal $k$.
The signal received at the BS, $\mat{Y} = [\vec{y}(1), \dots, \vec{y}(L)] \in \mathbb{C}^{M \times L}$, 
is obtained as
\be\label{received_signal}
\mat{Y} = \mat{H}\mat{P}\transpose+\mat{N},
\ee
where $\mat{H} = [\vec{h}_1, \dots, \vec{h}_{K}]$ is the column concatenation of the channel vectors from the $K$ terminals to the $M$ antennas at the BS, $\mat{P} = [\vec{p}_1, \dots, \vec{p}_{K}] \in \mathbb{C}^{L \times K}$ is the matrix containing the training sequences sent by the UTs, and $\mat{N}\in \mathbb{C}^{M \times L}$ represents additive noise.

By vectorizing the received signal in~(\ref{received_signal}), 
it can be expressed as
\be\label{rx_sigvec}
\mathrm{vec}(\mat{Y}) = \tilde{\mat{P}}\mathrm{vec}(\mat{H})+\mathrm{vec}(\mat{N}),
\ee
where $\tilde{\mat{P}} =(\mat{P} \otimes \mat{I}_{M})$.
Combining (\ref{ch_expr}) with~(\ref{rx_sigvec}) yields
\be \label{system_model}
\vec{y}=\mathrm{vec}(\mat{Y}) = \tilde{\mat{P}}\tilde{\mat{R}}\radixM \vec{\eta}+\mathrm{vec}(\mat{N}),
\ee
where 
\begin{displaymath}
\tilde{\mat{R}}\radixM = \left( \begin{array}{ccc}
\mat{R}_1\radixM & \dots & 0 \\
0 & \mat{R}_{k}\radixM & 0 \\
0 & \dots & \mat{R}_{K}\radixM  \end{array} \right),
\end{displaymath}
and $\vec{\eta}\in \mathbb{C}^{r}$ is the vector concatenation of the fast fading process coefficients, i.e. $\vec{\eta}^T=[\vec{\eta}_1^T,\ldots,\vec{\eta}_K^T]$ with $r =\sum_{i =1}^{K} r_i$. The objective of the receiver is to jointly estimate $\vec{h}_1 \ldots \vec{h}_K$ from the received signal $\mat{Y}$.

\section{Channel Identifiability}
\label{section_identifiability}

In this section, we establish bounds on the length of the probing sequence that is required to be able to identify the channel; to this aim, we focus on the noiseless case ($\mat{N}=\bf 0$), and establish for which pilot length $L$ the knowledge of $\mat{Y}$ is sufficient to uniquely identify $\mat{H}$. Note that because of \eqref{ch_expr} and since $\mat{R}_k\radixM$ is assumed full column rank, it is equivalent to identify $\vec{h}_k$ and $\vec{\eta}_k$. Furthermore, since in the noise-free case, \eqref{system_model} yields the trivial system of linear equations $\vec{y} = \tilde{\mat{P}}\tilde{\mat{R}}\radixM \vec{\eta}$, all the channel vectors $\vec{h}_1,\ldots, \vec{h}_K$ can be estimated from $\mat{Y}$ iff $\tilde{\mat{P}}\tilde{\mat{R}}\radixM$ has full column rank. Therefore, we will focus on the identifiability condition
\be \label{eq_identifiability}
\mathrm{rank}\left(\tilde{\mat{P}}\tilde{\mat{R}}\radixM\right) = r.
\ee

\subsection{General Case}

We first consider the general case where no particlar assumption is made on the relative position of the subspace spanned by the correlation matrices $\mat{R}_1\radixM,\ldots,\mat{R}_K\radixM$.
 Note that
\begin{eqnarray}
	\mathrm{rank}(\tilde{\mat{P}}\tilde{\mat{R}}\radixM) &\leq &\min\left(\mathrm{rank}(\tilde{\mat{P}}),\mathrm{rank}(\tilde{\mat{R}}\radixM)\right)\\
	&\leq & \min\left(M\cdot\mathrm{rank}(\mat{P}),r\right).\label{rank_ineq1}
\end{eqnarray}
The identifiability criterion \eqref{eq_identifiability} imposes to fulfill \eqref{rank_ineq1} with equality, which requires $M\cdot\mathrm{rank}(\mat{P})\geq r$.
Finally, since $\mat{P}$ has dimension ${L \times K}$, we have $L \geq \mathrm{rank}(\mat{P})$, which yields the necessary condition for identifiability
\be \label{eq_genreral_necessary}
L \geq \frac{r}{M}.
\ee

We now establish a sufficient condition. Using Sylvester's rank inequality, we obtain
\begin{eqnarray}
 \mathrm{rank}(\tilde{\mat{P}}\tilde{\mat{R}}\radixM) &\geq& \mathrm{rank}(\tilde{\mat{P}}) + \mathrm{rank}(\tilde{\mat{R}}\radixM)-KM \\
 &\geq& M \cdot\mathrm{rank}(\mat{P})+r-KM.  \label{eq_rankcondition1}
\end{eqnarray}
Note that taking
\be \label{eq_sufficientgeneral}
\mathrm{rank}(\mat{P})=K
\ee
in \eqref{eq_rankcondition1} yields $\mathrm{rank}(\tilde{\mat{P}}\tilde{\mat{R}}\radixM)\geq r$, i.e. it guarantees identifiability. Since $\mathrm{rank}(\mat{P})\leq \min(L,K)$, the sufficient condition \eqref{eq_sufficientgeneral} also imposes $L\geq K$, which is consistent with the fact that we do not exploit the spatial properties of the channel to reduce the training length, and therefore $L$ must be at least equal to the number of users $K$ to ensure that they can be distinguished. 

Note that the bounds on the training length $L$ obtained through the necessary condition \eqref{eq_genreral_necessary} and the inequality $L\geq K$ associated to the sufficient condition \eqref{eq_sufficientgeneral} are not equal.

\subsection{Mutually Orthogonal Channel Subspaces}
\label{sec_mutually_orthogonal}

Let us consider the case of mutually orthogonal channel covariance matrices, i.e. $\mathrm{tr}\left(\mat{R}_i \mat{R}_j\right)=0 \ \forall i\neq j$.
In that case, the necessary condition \eqref{eq_genreral_necessary} can be shown to be sufficient as well. First, let us consider the case $r\leq M$. \eqref{eq_genreral_necessary} indicates that a training sequence of length at least $L=1$ is required; we show that this is indeed sufficient. For this, consider the training sequence of user $k$, $\vec{p}_k = [p_k(1)]$ which is reduced to length 1, and assume that  $p_k(1)\neq 0 \ \forall k$.
It is trivial to see that 
\be
\tilde{\mat{P}}\tilde{\mat{R}}\radixM=\left(p_k(1)\mat{R}_1\radixM \ \ldots \  p_k(K)\mat{R}_K\radixM \right)
\ee
is full column rank thanks to the orthogonality of the column subspaces of $\mat{R}_1\radixM \ \ldots \ \mat{R}_K\radixM$, and therefore the identifiability condition \eqref{eq_identifiability} is fulfilled.
A similar result can be obtained for the case $r>M$, where it can be shown that a training sequence of length $L=\lceil \frac{r}{M}\rceil$ is sufficient to ensure identifiability.
We conclude that in the case of mutually orthogonal subspaces, the bound \eqref{eq_genreral_necessary} is tight.

Note that this case is of particular interest in Massive MIMO, since it shows that when $\frac{r}{M}$ is small, $L$ can be made small, and in particular channel identifiability can be achieved using pilot sequences of length $L<K$.

\subsection{Identical Channel Subspaces}

Let us now focus on the case where the channels of all the users live in the same linear subspace; this can be represented without loss of generality by taking $\mat{R}_1\radixM=\ldots=\mat{R}_K\radixM=\mat{R}\radixM$ for some full column rank matrix $\mat{R}\radixM \in \mathbb{C}^{M \times d}$.
In that case, we have $\tilde{\mat{P}}\tilde{\mat{R}}\radixM =\mat{P}\transpose \otimes  \mat{R}\radixM$, and
\be \label{eq_rank_identical}
\mathrm{rank}(\tilde{\mat{P}}\tilde{\mat{R}}\radixM) = 
\mathrm{rank}(\mat{P}) \cdot d.
\ee
Since $r=Kd$, combining \eqref{eq_identifiability} and \eqref{eq_rank_identical} yields the identifiability condition   $\mathrm{rank}(\mat{P}) = K$.
Note that this is the same condition as \eqref{eq_sufficientgeneral}, however in the case of identical subspaces across the users, it is both necessary and sufficient, while in the general case it is not necessary, as shown in Section \ref{sec_mutually_orthogonal}.


\section{Minimum length pilot sequence under estimation error constraint}
\label{section_optimization}

Let us now consider the more general setting where noise is present, and the covariance matrices of the users can be arbitrary; Linear Minimum Mean Square Error (LMMSE) estimation of the channels is assumed. In this context, we tackle the design of short pilot sequences under an estimation error constraint.

\subsection{Error covariance matrix}

The LMMSE estimator of the fast fading coefficients between all users and the BS array, $\hat{\vec{\eta}}$, is
\be
\hat{\vec{\eta}} = \mat{C}_{\vec{\eta}\vec{y}}\mat{C}_{\vec{y}}^{-1}\vec{y},
\ee
where  $\mat{C}_{\vec{\eta}\vec{y}} = \tilde{\mat{R}}\radixH \tilde{\mat{P}}\hermitian$, and $\mat{C}_{\vec{y}} = \tilde{\mat{P}} \tilde{\mat{R}} \tilde{\mat{P}}\hermitian + \sigma^2 \mat{I}_{LM}$.
The covariance matrix of the estimation error for $\vec{\eta}$ is given by~\cite{libro_linearestimation}
\bea
\mat{C}_{\bf e,\vec{\eta}} &=& \mathbb{E}[(\hat{\vec{\eta}}-\vec{\eta})(\hat{\vec{\eta}}-\vec{\eta})\hermitian]\\
&=& \mat{I}_r-\tilde{\mat{R}}\radixH \tilde{\mat{P}}\hermitian(\tilde{\mat{P}}\tilde{\mat{R}}\tilde{\mat{P}}\hermitian+\sigma^2\mat{I}_{LM})^{-1} \tilde{\mat{P}}\tilde{\mat{R}}\radixM.
\eea
Considering \eqref{ch_expr}, we define $\hat{\vec{h}} = \tilde{\mat{R}}\radixM \hat{\vec{\eta}}$. The covariance matrix of the estimation error on $\vec{h}$ is therefore
\bea
\mat{C}_{\bf e} &=&\mathbb{E}[(\hat{\vec{h}}-\vec{h})(\hat{\vec{h}}-\vec{h})\hermitian]\\
&=& \tilde{\mat{R}}\radixM \mat{C}_{\bf e,\vec{\eta}} \tilde{\mat{R}}\radixH \\
&=& \tilde{\mat{R}}-\tilde{\mat{R}} \tilde{\mat{P}}\hermitian(\tilde{\mat{P}}\tilde{\mat{R}}\tilde{\mat{P}}\hermitian+\sigma^2\mat{I}_{LM})^{-1} \tilde{\mat{P}}\tilde{\mat{R}}. \label{eq_Ce_expanded}
\eea
In order to control the accuracy of the channel estimation process across all the users, we assume that we wish to uniformly bound the estimation error on all dimensions of $\vec{h}$ by a given constant $\epsilon > 0$. This can be done by requiring that all the eigenvalues of $\mat{C}_{\bf e}$ are lower or equal to $\epsilon$, which we denote\footnote{For two positive semidefinite matrices $\mat{A}$ and $\mat{B}$, $\mat{A} \preceq \mat{B}$ is a shorthand notation for the condition that $\mat{B}-\mat{A}$ is positive semidefinite.} $\mat{C}_{\bf e} \preceq \epsilon\mat{I}$.

\subsection{Pilot length minimization algorithm}

We now formulate the pilot length minimization as a rank minimization problem over a convex set. Recall that we seek the minimum $L$ for which there exists a $L\times K$ matrix $\mat{P}$ that satisfies $\mat{C}_{\bf e}~\preceq~\epsilon\mat{I}$. Thus we seek to solve the optimization problem
\begin{eqnarray}\label{eq:rank}
	& \underset{ \mat{P}\in \mathbb{C}^{L \times K}} \min & L\\
	& \mathrm{s.t. } &  \mat{C}_{\bf e} \preceq \epsilon\mat{I}. \nonumber
\end{eqnarray}
Letting $\mat{X}= \mat{P}\hermitian\mat{P} \in \mathbb{C}^{K\times K}$, minimizing $L$ is equivalent to minimizing $\rank(\mat{X})$, i.e. \eqref{eq:rank} becomes:
	\bea \label{eq_sdp1}
	&\underset{\mat{X} \succeq {\bf 0}} \min & \rank(\mat{X})\\
	&\mathrm{s.t. } &\mat{C}_{\bf e} \preceq \epsilon\mat{I}. \nonumber
	\eea	
This problem can be solved efficiently, but approximately, through heuristics. Following \cite{Fazel04rankminimization}, we consider a regularized smooth surrogate of the rank function, namely $\log\det(\mat{X}+\delta \mat{I})$ for some small $\delta$, which yields:
	\bea \label{eq_concave_objective}
	&\underset{\mat{X} \succeq {\bf 0}} \min & \log\det(\mat{X}+\delta \mat{I})\\
	&\mathrm{s.t. } &\mat{C}_{\bf e} \preceq \epsilon\mat{I}. \nonumber
	\eea

The objective function in \eqref{eq_concave_objective} is concave, however it is smooth on the positive definite cone; a possible way to approximately solve this problem is to iteratively minimize a locally linearized version of the objective function, i.e. solve
\begin{eqnarray}  \label{eq:rank_minimization2}  
				\mat{X}_{t+1} = \arg &  \underset{\mat{X} \succeq {\bf 0}} {\min } & \mathrm{Tr}(\mat{X}_t+\delta \mat{I})^{-1}\mat{X}\\
				& \mathrm{s.t. } & \mat{C}_{\bf e} \preceq \epsilon\mat{I}. \nonumber
\end{eqnarray}
until convergence to some $\bar{\mat{X}}$. We suggest to initialize the algorithm by choosing $\mat{X}_0$ as the rank-1, all-ones matrix $\mathbf{1}_{K\times K}$.\\

Let us now focus on the constraint $\mat{C}_{\bf e} \preceq \epsilon\mat{I}$. Decomposing $\tilde{\mat{R}}\radixM = \mat{U}\mat{\Lambda}^{1/2}$, where $\mat{\Lambda}$ and $\mat{U}$ are respectively the diagonal matrix containing the eigenvalues of $\tilde{\mat{R}}$ and the associated eigenvectors, and using \eqref{eq_Ce_expanded}, we obtain
\begin{equation} \label{eq_constraint_maxerror}
\mat{C}_{\bf e} \preceq \epsilon\mat{I} \ \Leftrightarrow \ \tilde{\mat{R}}\radixH \left(\mat{X}\otimes\mat{I}_M\right)\tilde{\mat{R}}\radixM \succeq \left(\epsilon^{-1}\mat{\Lambda}- \mat{I}_r\right) \sigma^2.
\end{equation}
Note that since this constraint is convex, \eqref{eq:rank_minimization2} is a convex optimization problem that can be efficiently solved numerically.\\

Due to the various approximations involved in transforming \eqref{eq_sdp1} into \eqref{eq:rank_minimization2}, $\bar{\mat{X}}$ might not be strictly rank-deficient, but it can have some very small eigenvalues instead. It is therefore necessary to apply some thresholding on these eigenvalues to recover a stricly rank-deficient solution. Let us denote by $\vec{e}_k$ the eigenvector associated to the $k$-th eigenvalue $v_k$ of $\bar{\mat{X}}$, $k=1\ldots K$, with $v_1\geq \ldots \geq v_K\geq 0$. We then let 
\begin{eqnarray} \label{eq_eigthresholding}
L&=& \max_{i=1\ldots K} i\\
&& \mathrm{s.t.} \quad v_i \geq \epsilon_s \nonumber
\end{eqnarray}
for a suitably chosen (small) $\epsilon_s$, and obtain the matrix of optimized training sequences as $\bar{\mat{P}} = [\vec{e}_1, \dots, \vec{e}_{L}]^T$.\\

The proposed algorithm is shown in Algorithm~\ref{algo:algo1}.

	\begin{algorithm}
		\caption{Minimum length pilot sequence computation with estimation error constraint.}
		\label{algo:algo1} 
		\begin{algorithmic}
			\STATE{Initialize $\mat{X}_0 \leftarrow \mathbf{1}_{K\times K}$.}
			\REPEAT
			\STATE{$\begin{array}{rcl} 
			\mat{X}_{t+1} &\leftarrow& \arg   \underset{\mat{X} \succeq {\bf 0}} {\min }  \mathrm{Tr}(\mat{X}_t+\delta \mat{I})^{-1}\mat{X}\\
				& \mathrm{s.t. }&  \tilde{\mat{R}}^{H/2} \left(\mat{X}\otimes\mat{I}_M\right)\tilde{\mat{R}}\radixM \succeq \left(\epsilon^{-1}\mat{\Lambda}- \mat{I}_r\right) \sigma^2.\end{array}$ }
			\UNTIL{convergence to $\bar{\mat{X}}$.}
			\STATE{Compute $L$ according to \eqref{eq_eigthresholding}.}
			\STATE{\textbf{Output:} $\bar{\mat{P}} = [\vec{e}_1, \dots, \vec{e}_{L}]^T$.}
		\end{algorithmic}
	\end{algorithm}

\section{Numerical results}
\label{section_simulations}

Algorithm~\ref{algo:algo1} has been bechmarked numerically. For each realization of the covariance matrices, we applied the proposed algorithm to compute $\bar{\mat{P}}$ (the solution to \eqref{eq:rank_minimization2} was obtained via the numerical solver CVX \cite{cvx}, and parameter $\delta$ was set to $10^{-4}$). We note that although we do not provide any convergence proof, the proposed method has demonstrated reliable convergence in the simulations. Note also that we can not claim any global optimality for the obtained solution $\bar{\mat{X}}$, and indeed simulations have shown that the convergence point depends on the initialization, with the rank-1 initialization $\mat{X}_0=\mathbf{1}_{K\times K}$ giving the best results.\\
	
\begin{figure}[htb]
\centering
  \includegraphics[scale=0.6]{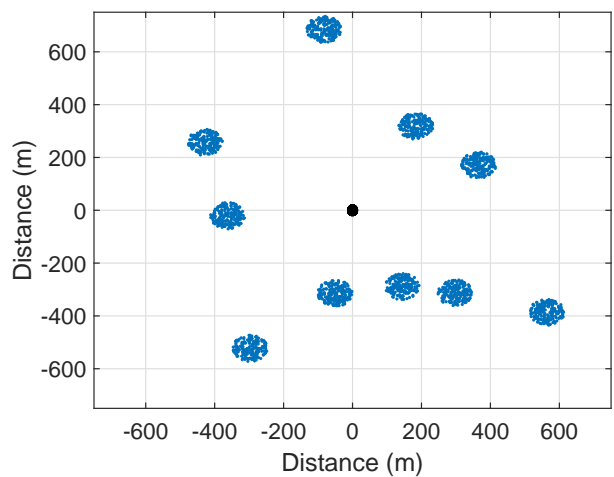}\caption{Example realization of the system geometry used to compute the channel covariance matrices. $K=10$ users are randomly spread between 250m and 750m of the BS, materialized by the UCA at the origin. Each user is surrounded by 200 scatterers randomly spread on a disc of radius 50m.}
	\label{fig:geometry}
\end{figure}

The scenario considered in this section is that of a uniform circular array (UCA) of diameter $2$m, consisting of $M \in \{16, 30\}$ antenna elements (AE), which serves $K \in \{5, 10\}$ UTs randomly distributed around the BS at a distance between $250$ and $750$m.  We assume that $200$ scatterers distributed randomly on a disc of radius $50$m centered on each terminal are causing fast fading (see Fig.~\ref{fig:geometry}).
The covariance matrices are generated by a ray-tracing procedure based on the one-ring channel model \cite{OneRingModel}, with a central frequency of $1.8$~GHz, and are normalized such that $\mathrm{trace}(\mat{R}_k)=...=\mathrm{trace}(\mat{R}_K)=1$ (this can be achieved in reality via power control). According to this model, the support of the angle of arrivals associated to a given UT is limited,  which yields covariance matrices with few large eigenvalues. We have applied a threshold to these eigenvalues to obtain the ranks $r_k$ that ensure that at least 99\% of the energy of the full-rank matrix is captured by the rank-deficient model. The noise variance at each BS antenna element is chosen as $\sigma^2 = 10^{-4}$.\\

\begin{figure}[thb]
\centering
		\includegraphics[scale =0.6]{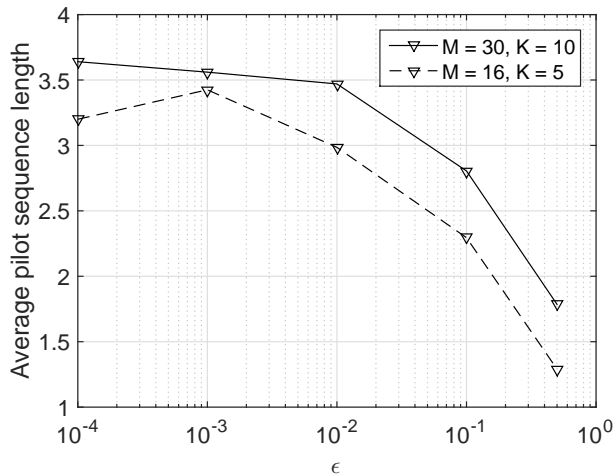}\caption{Average length $L$ of the pilot sequences as a function of the error threshold $\epsilon$ for  $(M,K)= (30,10)$ (solid line) and  $(16,5)$ (dashed).}\label{fig:avgL}
	\end{figure}

Fig.~\ref{fig:avgL} shows the average length of the pilot sequences obtained by the proposed algorithm over $200$ realizations as a function of the error threshold $\epsilon$. Clearly, $L$ decreases when the constraint on the maximum estimation error is relaxed (for large $\epsilon$). In all scenarios, the proposed algorithm yields pilot sequences having an average $L$ significantly lower than $K$. Note also that this gain appears to be more pronounced for larger antenna arrays.
\begin{figure}
\centering
	\includegraphics[scale=0.6]{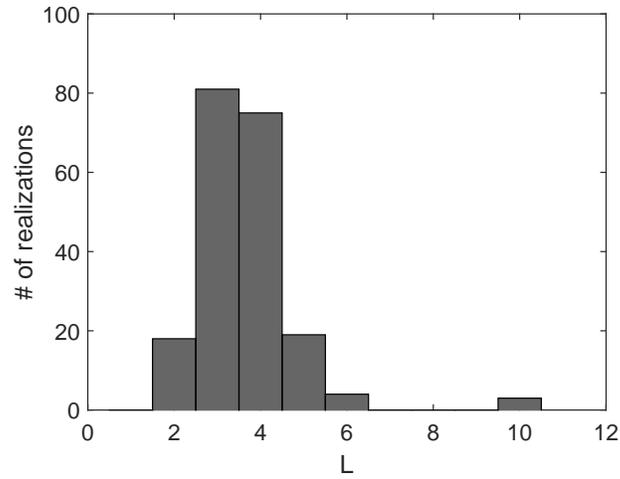}\caption{Histogram of $L$ over $200$ realizations of the covariance matrices. $(M,K)=(30,10)$, and  $\epsilon = 10^{-4}$.}
	\label{fig:histoL}
\end{figure}

Fig.~\ref{fig:histoL} depicts the histogram (over $200$ realizations) of $L$ in the scenario with $M=30$ AE and $K=10$ UTs, when the constraint on the maximum error is set to $\epsilon=10^{-4}$. In most of the realizations, the length of the pilot sequence is significantly smaller than the number of users. 

\begin{figure}[htb]
	\centering
		\includegraphics[scale=0.6]{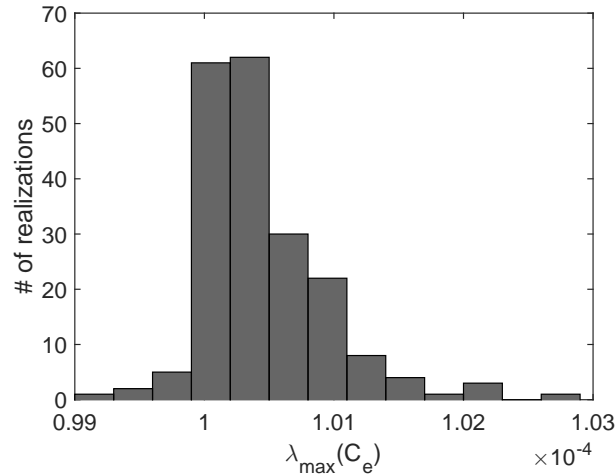}\caption{Histogram of the maximum eigenvalue of $\mat{C}_{\bf e}$ achieved by using the pilot sequences $\bar{\mat{P}}$, for the target $\epsilon = 10^{-4}$. $(M,K)=(30,10)$.}\label{fig:histoLambda}
	\end{figure}

Note that because of the thresholding procedure described in \eqref{eq_eigthresholding}, $\mat{X}= \bar{\mat{P}}\hermitian\bar{\mat{P}}$ does not necessarily satisfy \eqref{eq_constraint_maxerror} exactly, although $\bar{\mat{X}}$ does. In order to verify that the maximum estimation error constraint is still approximately satisfied, we compute the maximum eigenvalue of $\mat{C}_{\bf e}$ for the length-$L$ pilot sequence obtained after thresholding. The results, in Fig.~\ref{fig:histoLambda} for $(M,K)=(30,10)$, show that this error is not significantly greater than the target $\epsilon$.


\section{Conclusions}

We have analyzed how channel correlation and the knowledge of per-user covariance matrices can significantly reduce the duration of channel estimation in Massive MIMO systems. We have established bounds on the duration of estimation based on the extreme cases of identical and orthogonal channel covariance matrices. For the general case, we introduced an algorithm to design short training sequences for a given set of user covariance matrices. The obtained pilot sequences are in general non-orthogonal, however they ensure that the channel estimation error variance is uniformly upper-bounded by a chosen constant over all channel dimensions.
Simulations have shown that the proposed technique can yield pilot sequences of length significantly smaller than the number of users in the system.

\newpage

\bibliographystyle{IEEEtran}
\bibliography{IEEEabrv,refen}
\end{document}